# Development and application of an optical TPC for charged particle track structure imaging in microdosimetry

U. Titt, V. Dangendorf*, B. Großwendt, H. Schuhmacher

*Physikalisch-Technische Bundesanstalt,  Bundesallee 100, 38116 Braunschweig, Germany*

**Abstract:**

An imaging system for measuring the track structure of charged particles in a low-pressure gas is presented. We describe the experimental method and compare experimental results with first results of a simulation of the chamber response.

**\* Corresponding author:**
Tel: +49 531 592 7525; fax: +49 531 592 7015; e-mail: volker.dangendorf@ptb.de

## 1. Introduction

Radiation action in matter and namely in living tissue depends on the microscopic details of energy transfer in the micrometer and nanometer range. Since it is not possible to measure with such a high resolution in condensed phase, various methods have been developed to substitute this measurements a) by measurements in low-pressure gases or b) by simulating the radiation transport through matter by Monte Carlo (MC) models and calculating the deposited energy or the number of ionization events in the volumes of interest.

Most of the experimental approaches utilise small gas-filled cavities (tissue-equivalent proportional counters, TEPCs) whose geometrical size, scaled by the density ratio between tissue and gas simulates the size of the biological entity of interest. These methods are of course restricted to detector sizes a few mm in diameter at a pressure of a few hPa, which correspond to tissue volumes of a few 100 nm at best. On the other hand, the sizes of the radiation-sensitive cell structures range from several 100 nm (chromosome), 30 nm (diameter of chromatine fibre) and 10 nm (nucleosome) down to 2 nm (distance of the DNA helices). An adequate instrument therefore should cover the full range of these sizes.

On the other hand, an MC model appropriate for this purpose must simulate not only the degradation of primary particles but also the production and propagation of secondary and higher-order particles, for instance, low-energy electrons in the case of ionization processes or heavy charged particles in the case of neutron interactions. A prerequisite for such a Monte Carlo model is a comprehensive set of interaction cross sections with respect to the essential effects governing particle degradation, combined with a benchmarking of Monte Carlo results with appropriate experimental data.

In this contribution we describe a gas-filled imaging system which measures the spatial pattern of energy deposition in a simulated cavity a few micrometer in diameter with a potential resolution of several 10 nm. With this instrument we obtain the full statistical correlation of ionisation events from primary and higher-order charged particles in this volume scale whereas a TEPC provides only integral information about the total energy deposited in a fixed volume. Furthermore, it provides a benchmark for results of MC calculations. Here we compare results of particle track structure calculations with the experimental results obtained with the detector.

**2. The Detector**

The experimental method is based on a time projection chamber with a parallel drift field, parallel-plate charge and light amplification steps and optical readout (Optical Avalanche Chamber, OPAC). Fig1 shows the experimental set-up. A detailed description of the instrument, its readout facilities and the basic properties can be found in [1]. The chamber is operated with triethylamine (TEA) vapour at a pressure of 10 hPa. At that pressure a distance of 1 mm in the gas corresponds to 40 nm in tissue. TEA is a very efficient scintillator in the near UV, and its atomic composition is close to that of tissue.

In a previous paper [1] we described the basic properties of the gas amplification and scintillation processes in low-pressure TEA. As an example of the images obtained with the OPAC, Fig. 2 shows the tracks of a proton of 5 MeV energy and an electron of a few keV energy. More track samples of various ions and electrons are presented in recent papers [1, 2], in which we also reported on results of particle identification by analysing the ionisation profile of the tracks [2].

## 3. Response calculations

Here we report on response calculations of the chamber by simulation of the charged particle transport through the gas and the transport of the fast and slow electrons through the gas.

The transport of the ions in TEA is calculated in continuous slowing-down approximation (CSDA) based on Ziegler *et al.* [3]. The electronic stopping fraction of the energy loss is converted into δ-electrons, for which the energy distribution is derived from the Rudd model [4]. Fig. 3 shows the δ-electron energy distribution for an α-particle and a proton, which have the same energy loss (d$E$/d$x$) in the gas. While the total energy transmitted to electron is the same, the fraction of higher energetic electrons (keV-electrons) is greater for the α-particles. This is due to fact that the speed of the heavier particle is about 2,5 times higher than that of the proton.

The initial angular distribution of the electrons is roughly approximated by the binary encounter (BEA) theory (see, e.g,. [4]). However, since the electron trajectory is soon randomised by large angle scattering, the initial emission angle is of only minor influence on the lateral extension of the particle track. For the transport of the emerging electrons, the cross sections for elastic scattering, excitation and ionisation are required. None of these is available for TEA. Therefore, as a first approximation, the cross sections for $CH_4$ from Großwendt [5] are used. The electron scattering angles are calculated by the approximation of Berger [6].The electrons are tracked till they reach subionisation energies. Then they are assumed to be at rest; their positions are recorded and form the undiffused ionisation track of the particle.

In the further simulation, the subionisation electrons are transported by the parallel electric field through the gas, taking the measured transversal and longitudinal diffusion into account [1]. The influence of the optical system is negligible in the central part of the image and therefore not included in the simulation. The distortions near the edges are neglected, because for the comparison only the central part is used. The optical image of the ionisation pattern which is formed in the amplification stages is therefore taken as the simulated response of the whole system and compared with the measured data.

Fig. 4 shows the results for the two particles whose electron distributions are shown in Fig.3. Even in the diffused images, the larger fluctuations due to the fewer but higher-energy $\delta$-electrons for the faster $\alpha$-particle are seen.

Fig. 4 suggests that an analysis of the cluster size distributions may be a suitable way of ion identification or mass separation for particles of same d$E$/d$x$. However, in a previous paper, we already discussed the possibility of particle identification due to this difference in average electron energy, because the faster electron transports a larger fraction of energy farther away from the central track core. The calculation of the transversal ionisation profile from the MC data shows the difference in the lateral extension of the ionisation track between two particles (see Fig.5) and agrees with the earlier experimental results (see [2])..

## 4. Conclusion

We described a particle track chamber with a high spatial resolution for application in radiation protection and radiation biology. We presented first results of track structure simulations which where folded with the chamber response (diffusion). The calculations confirm earlier results of particle identification based on d$E$/d$x$ and transversal ionisation density measurements.

**Acknowledgement**

We gratefully acknowledge the contributions of K. Tittelmeier, D. Mugai, W. Wendt and W. Heinemann to this work.

The work was supported by the EU, Contract No. FI4P-CT95-0024.

**Figure captions:**

Fig. 1: Schematic view of the optical TPC

Fig. 2: Track images of a proton of 5 MeV energy with an emerging fast δ-electron (a) and an electron of a few keV initial energy (b) are shown. The gas amplification in (b) is larger, therefore the image saturation there is higher despite the lower ionisation density.

Fig. 3: Energy distribution of the δ-electrons produced by protons and α-particles of 1 and 24 MeV energy. The ions have the same specific energy loss (dE/dx) in the gas and transfer approximately the same total energy to δ-electrons.

Fig. 4: Calculated response of the detector to a proton and α-particle of same dE/dx (see Fig. 3)

Fig 5: The difference in the velocity of the two particles of same d$E$/d$x$ yields different lateral ionization distributions.

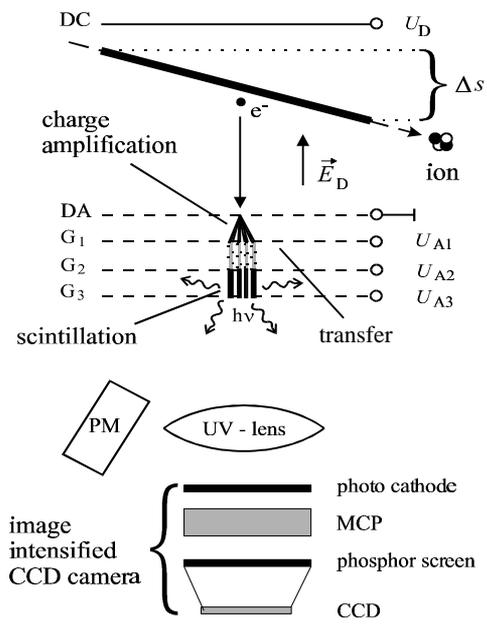

Fig.1

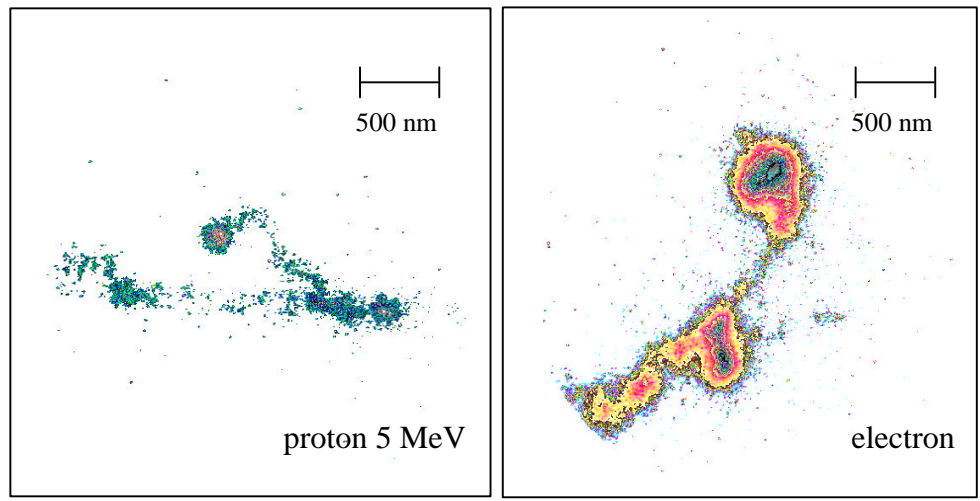

Fig.2

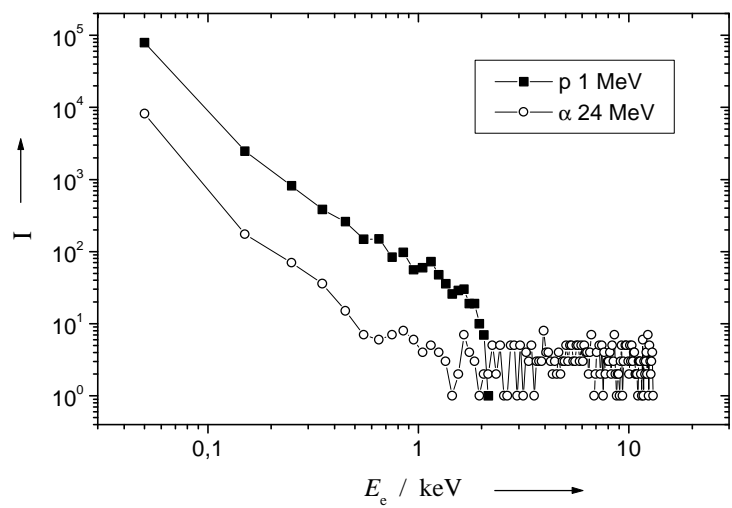

Fig. 3

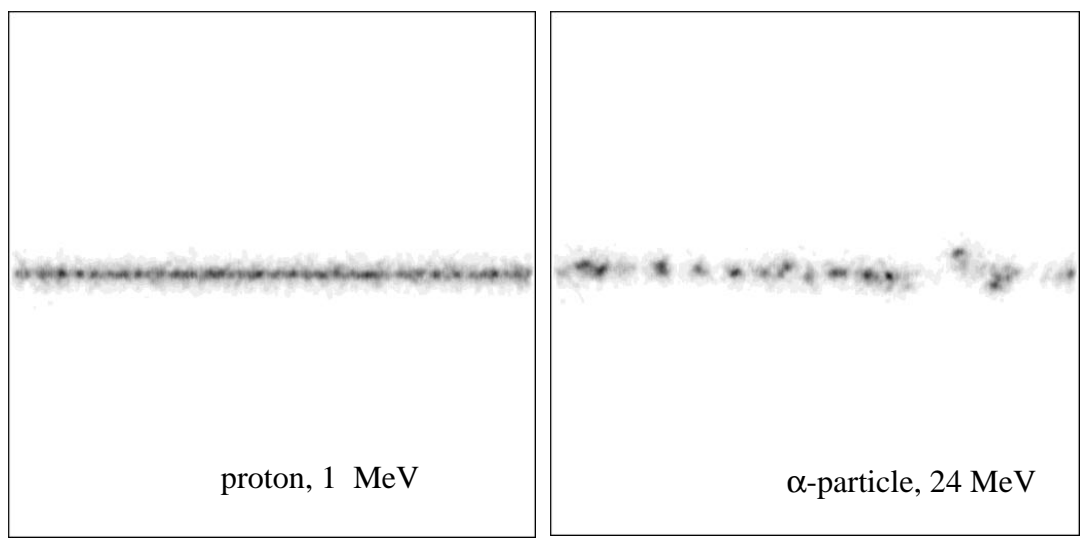

Fig.4

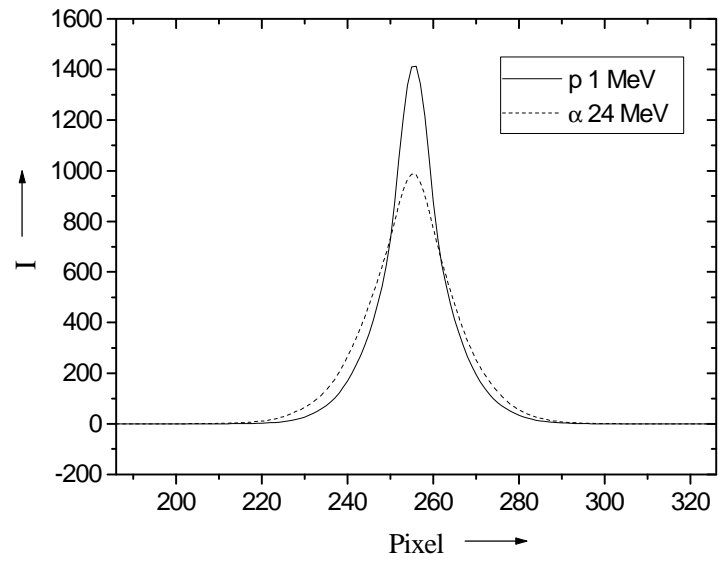

Fig. 5